\documentclass[twocolumn]{aastex61}
\usepackage{amsmath}
\usepackage{graphicx}
\usepackage{soul}

\newcommand{\gerg}{Gerg s$^{-1}$ cm$^{-2}$}
\newcommand{\ratio}{Z{:}H}

\begin{document}
\title{Connecting Giant Planet Atmosphere and Interior Modeling: Constraints on Atmospheric Metal Enrichment}
\author{Daniel Thorngren}
\affil{Department of Physics, University of California, Santa Cruz}
\author{Jonathan J. Fortney}
\affil{Department of Astronomy and Astrophysics, University of California, Santa Cruz}

\begin{abstract}
Atmospheric characterization through spectroscopic analysis, an essential tool of modern exoplanet science, can benefit significantly from the context provided by interior structure models.  In particular, the planet's bulk metallicity $Z_p$ places an upper limit on potential atmospheric metallicity.  Here we construct interior structure models to derive $Z_p$ and atmospheric metallicity upper limits for 403 known transiting giant exoplanets.  These limits are low enough that they can usefully inform atmosphere models.  Additionally, we argue that comparing $Z_p$ to the observed atmospheric metallicity gives a useful measure of how well-mixed metals are within the planet.  This represents a new avenue for learning about planetary interiors.  To aid in the future characterization of new planet discoveries we derive analytic prior predictions of atmosphere metallicity as a function of planet mass, and evaluate the effectiveness of our approach on Jupiter and Saturn. We include log-linear fits for approximating the metallicities of planets not in our catalog.
\end{abstract}

\section{Introduction}\label{sec:introduction}
Spectroscopic characterization of exoplanet atmospheres has proven to be an invaluable tool in understanding the nature and formation of giant planets.  Under the core accretion model of giant planet formation \citep[see][]{Pollack1996}, these planets are records of the disks from which they formed.  For example, the C/O ratio of a planet may depend on where it formed relative to the ice lines of water, methane, CO, and CO$_2$ and the relative accretion of solids and gas \citep{Oberg2011a,Madhusudhan2014,Mordasini2016,Espinoza2017}.  Many studies have collected emission and transmission spectra for purpose of determining molecular abundances, e.g. \cite{Swain2010, Line2014, Kreidberg2018, Wakeford2018}, often using the \emph{Spitzer} and/or \emph{Hubble} Space Telescopes.  These observations can also reveal the presence of hazes and clouds \citep[e.g.][]{Fortney2005,Sing2011, Gibson2012, Mandell2013, Morley2013, Kreidberg2014a} as well as atmospheric temperature structure, including whether a temperature inversion is present \citep{Knutson2008, Fortney2008, Burrows2008, Madhusudhan2011, Evans2016}.

With the recent successful launch of the Transiting Exoplanet Survey Satellite \citep[\emph{TESS}; see][]{Ricker2015}, many more planets amenable to spectroscopic follow-up are likely to be discovered \citep{Barclay2018,Sullivan2015,Huang2018}.  Additionally, the \emph{James Webb Space Telescope} \citep[\emph{JWST}; see][]{Gardner2006} will allow for measurements in new wavelength ranges with unprecedented precision \citep{Beichman2014,Bean2018}.

An important driver in atmospheric measurements is determining the metallicity of the planetary atmosphere \citep{Fortney2013}, which can be compared to predictions of formation models.  However due to degeneracies in determining atmospheric abundances \citep[first identified in][]{Benneke2012}, error bars on the abundances of atoms and molecules of interest can often be large \citep[see also][]{Griffith2014a,Line2016,Heng2017,Fisher2018}.  This can manifest itself as a strong prior dependence \citep[see e.g.][]{Oreshenko2017}.  As such, it would be helpful to have an additional source of information or constraint about the atmosphere's metallicity.

Interior structure models can help in this case.  For planets with known masses and radii, we can infer the bulk metallicity $Z_p$ through the use of planet evolution models which are used to understand the planetary radius over time, as in \cite{Thorngren2016}.  The equations of state for the most common metals (say, oxygen and carbon) at megabar pressures are similar enough \citep[e.g. compare][]{Thompson1990,French2009} that this approach is insensitive to the exact metals in question.  Iron's high density makes it an exception, but its lower abundance \citep{Asplund2009} makes this unimportant for our purposes.

Of course even knowing $Z_p$ exactly does not directly imply an atmosphere metallicity.  Even in the simplest model where the atmosphere and the entire H/He envelope share the same composition, some metals will likely be sequestered in the core.  In more complex models, interior composition gradients could lead to an increasing metallicity with depth in the H/He envelope \citep[e.g.][]{Leconte2012,Vazan2016}.  However, cases where $Z$ increases going outward in the planet will not be long-lived, succumbing either to Rayleigh-Taylor instability or ordinary convection.  Therefore the planet's bulk metallicity serves as an upper limit on the atmospheric metallicity.  We define the ``visible metal fraction $f$'' -- that observed in the atmosphere -- as the ratio of atmospheric metallicity $Z$ to the bulk metal, $Z_p$:
\begin{equation}
    Z = f Z_p
\end{equation}
The atmosphere cannot be more metal-rich than the interior, so $0 \leq f \leq 1$, and $Z_p$ is an \emph{upper limit} on the metallicity of the atmosphere.

Using this approach, we have previously helped to constrain metallicity in retrievals for two cases already.  For GJ 436b \citep{Morley2017}, interior structure models were helpful in contextualizing the inferred high atmospheric metallicity and connecting it to the large intrinsic flux suggested by the spectrum.  For WASP-107b \citep{Kreidberg2018}, we were able to help rule out a high-metallicity atmosphere, in agreement with the spectroscopic observations.

In this work, we seek to provide upper limits on atmospheric metallicity to assist with atmospheric retrieval modeling for every planet with sufficient data to support this.  We also discuss prior predictive distributions for the atmospheric metallicity, as well as fits to the upper limits so that future planet discoveries can easily produce limit estimates for planets with measured masses and radii.

Our data consists of transiting planets with RV and/or TTV follow up, downloaded and merged from the NASA Exoplanet Archive \citep{Akeson2013} and Exoplanets.eu \citep{Schneider2011}.  We consider only planets nominally massed between $20~M_\oplus$ and $20~M_J$ whose relative mass and radius uncertainties are $<50\%$.  We exclude hot Saturns, which we define as planets with mass $M < .5 M_J$ and flux $F > 0.5$ \gerg, as these planets are not well modeled by the inflated radius fits of \cite{Thorngren2018}.  An exception was made to include the potential JWST GTO object WASP-52 b, which was just over the line ($M = .46~M_J$, $F =~.65$ \gerg) but appears to be well-modeled.  These criteria resulted in the selection of 403 planets: 70 Saturns, 35 cool Jupiters, and 298 hot Jupiters.  The boundary between cool and hot Jupiters, by our definition, is .2 \gerg, below which significant radius inflation does not occur \citep{Miller2011,Demory2011}.

\section{Methods}\label{sec:methods}
Following \cite{Fortney2013}, consider a mass $M$ of gas with a metal mass fraction $Z$.  The mass of the hydrogen and helium is $M (1-Z)$, and the mass of the metals (everything else) is $MZ$.  Thus, given the mean molecular mass of the hydrogen ($\mu_H$) and metals ($\mu_Z$), the number of hydrogen and metal molecules is $N_H = M (1-Z) (X / (X+Y)) / \mu_H$ and $N_Z = M Z / \mu_Z$ respectively.  From this, we can compute the metal abundance ratio $\ratio$ (by number) as:
\begin{align}
    \ratio = \frac{N_Z}{N_H} &= \frac{M Z / \mu_Z}{M (1-Z)(X/(X+Y))/\mu_H} \\
    &= \frac{1+Y/X}{(Z^{-1} - 1)(\mu_Z/\mu_H)}\label{eq:nz}
\end{align}

Satisfyingly, this is independent of mass and only depends on $Z$, the H/He mass ratio $Y/X$, and the ratio of the mean molecular masses $\mu_Z/\mu_H$.  For our calculations, we use $\mu_H = 2 $ AMU (molecular hydrogen), $\mu_Z=18$ AMU (water), and $Y/X = .3383$ \citep{Asplund2009}.  Models reflecting individual planetary chemistry can be similarly constructed; as \cite{Heng2018} reminds, ``atmosphere metallicity'' is ambiguous, so extra care should be taken here.  Often in atmosphere modeling, this is parameterized in units relative to the metal abundance of the solar photosphere $\ratio_\odot = 1.04 \times 10^{-3}$ \citep{Asplund2009}.  We will use these units implicitly for the remainder of this letter.

In some atmospheric retrievals, the authors have opted not to lock different metal abundances to fixed ratios \citep[e.g.][]{Oreshenko2017}.  For these cases Eq. \ref{eq:nz} can still be useful.  To handle this, one must compute the total metallicity from the individual abundances (potentially making assumptions about unmodelled abundances).  One should also compute the mean molecular mass of the metals if they differ significantly from the assumed 18.  Using the new mean molecular mass of the metals $\mu_Z$, our tabulated $\ratio$ can simply be scaled by a factor of $18/\mu_Z$.  Note that this procedure only informs us of the total metal abundance, not individual molecular abundances.

From here we can proceed in two different ways.  First, in \S \ref{sec:priorPredictive}, we will combine Eq. \ref{eq:nz} with the mass metallicity relation from \cite{Thorngren2016}.  This results in a distribution for $\ratio$ which depends only on $f$ and the planet mass.  This is a useful as a baseline expectation for the planet population, but when considering individual planets we wish to also account for their observed radii, insolation, and age to get a more precise estimate.  For this, we combine Bayesian statistical models with interior structure models in \S \ref{sec:statistics}, which we then apply separately to each planet from our sample in turn.  The results of these calculation are discussed in \S\ref{sec:results}.

\subsection{Prior Predictive}\label{sec:priorPredictive}
A simple way to estimate the bulk metallicity of the planet is to make use of the planetary mass-metallicity relation we identified in \cite{Thorngren2016}, which takes the following form:
\begin{align}
    M_Z = \alpha' M^{\beta'} 10^{\pm \sigma_Z}
\end{align}
When $M_Z$ and $M$ are in Jupiter masses, $\alpha'=.182$, $\beta'=.61$, and $\sigma_Z=.26$.  We can neglect uncertainty in the parameters because the predictive uncertainty is dominated by the residual spread $\sigma_Z$.  This can be easily converted to a prior on bulk $Z_p$ as follows:
\begin{align}
    \log(M_Z) &= \alpha + \beta' \log(M) \pm \sigma_Z \\
    \log(M_Z/M) &= \alpha + (\beta'-1) \log(M) \pm \sigma_Z \\
    \log(Z_p) &= \alpha + \beta \log(M) \pm \sigma_Z \label{eq:mzRelation}
\end{align}
Here, $\alpha = \log_{10}(\alpha') = -.7395$ and $\beta = \beta' - 1 = -.39$ for brevity.  Combining equations \ref{eq:nz} and \ref{eq:mzRelation}, we can produce a prior on the relative number fraction of metals:
\begin{align}
    \ratio = \frac{1+Y/X}{(\mu_Z/\mu_H)(f^{-1}10^{-\alpha \pm \sigma_Z} M^{-\beta}-1)}\label{eq:priorPredictive}
\end{align}

From this, we can compute the expected amount of metal in an atmosphere given the mass of the planet and $f$.  The maximum atmospheric metal abundance $\ratio_\text{max}$ occurs when $f = 1$. To account for the additional information available from radius, age, and flux, we will include structure evolution modeling using a Bayesian framework in the next section.  These techniques are not wholely separate, however: Eq. \ref{eq:priorPredictive} is the prior predictive distribution with respect to that more sophisticated model.

\subsection{Statistical Models}\label{sec:statistics}
Our statistical model seeks to identify structure parameters which reproduce the observed radius $R_\mathrm{obs}$, accounting for the observational uncertainty $\sigma_R$.  The parameters we consider are the planet mass $M$ in Jupiter masses, the bulk planet metallicity $Z_p$, the anomalous heating efficiency $\epsilon$, and the age of the planet $t$ in Gyr.  Thus, we construct the following likelihood:
\begin{align}
    p(R_\mathrm{obs}|M,Z_p,\epsilon,t,\sigma_R) &= \label{eq:likelihood} \\
    \mathcal{N}(R_\mathrm{obs}|&R(M,Z_p,\epsilon,t),\sigma_R) \nonumber
\end{align}
Here, $R(M,Z_p,\epsilon,t)$ refers to the radius output of our structure models, and $\mathcal{N}(\mu,\sigma)$ is the a normal distribution with mean $\mu$ and standard deviation $\sigma$; $\mathcal{N}(x|\mu,\sigma)$ indicates the distribution's PDF should be evaluated at $x$ (the same notation will be used for other distributions later).

The priors for $M$ and $t$ are the observed mass and age of the planet, with the latter truncated between 0 and 14 Gyr, since we are confident that the planets are not older than the universe.
\begin{align}
    p(M) &\sim \mathcal{N}(M_{obs},\sigma_M) \label{eq:firstprior} \\
    p(t) &\sim \mathcal{TN}(t_{obs},\sigma_t,0,14)
\end{align}
We use $\mathcal{TN}(\mu,\sigma,x_0,x_1)$ to refer to a truncated normal distribution with mean $\mu$, standard deviation $\sigma$, and upper and lower limits $x_0$, $x_1$.  The prior for $Z_p$ comes from the mass-metallicity relation (Eq. \ref{eq:mzRelation}).
\begin{align}
    p(Z_p|M) &\sim \mathcal{LN}(\alpha+\beta \log_{10}(M),\sigma_Z) \label{eq:zprior}
\end{align}
We use $\mathcal{LN}(\mu,\sigma)$ to indicate a log normal distribution, where the $\log_{10}$ of the parameter is normally distributed with mean $\mu$ and standard deviation $\sigma$.

Hot Jupiter radius inflation represents a complicating factor in constructing evolution models for these objects.  We handle the anomalous heating efficiency $\epsilon$ using the Gaussian process posterior predictive results from \cite{Thorngren2018}.  There we inferred anomalous heating as a function of flux by adjusting it to match the modelled radius to the observed radius.  The composition was assumed to follow the same distribution as the warm giant planets, since they are in similar mass and orbital regimes.  Because of their extra degree of freedom, we see larger (but manageable) uncertainties on the bulk metallicities for hot Jupiters.
\begin{align}
    p(\epsilon) &\sim \mathcal{LN}(\epsilon(F),\sigma_{\epsilon}(F)) \label{eq:lastprior}
\end{align}

Thus, we are using the trends in composition and heating efficiency $\epsilon$ that reproduced observed radii of the giant planet population as the priors for individual planets.  Combining the likelihood (Eq. \ref{eq:likelihood}) and priors (Eq. \ref{eq:firstprior}- \ref{eq:lastprior}), we obtain a posterior proportional to:
\begin{align}
    p(&M,Z_p,\epsilon,t|R_{obs},\sigma_R) \propto \\
    &p(R_\mathrm{o}|M,Z_p,\epsilon,t,\sigma_R) p(M)
        p(Z_p|M) p(\epsilon) p(t) \nonumber \\
    \propto &\mathcal{N}(R_\mathrm{obs}|R(M,Z_p,\epsilon,t),\sigma_R)
        \mathcal{N}(M|M_{obs},\sigma_M) \label{eq:posterior}\\
    & \mathcal{LN}(Z_p|\alpha+\beta \log_{10}(M),\sigma_M) \nonumber
    \mathcal{LN}(\epsilon|\epsilon(F),\sigma_{\epsilon}(F)) \nonumber \\
    & \mathcal{TN}(t|t_{obs},\sigma_t,0,14) \nonumber
\end{align}

\begin{figure}[b!]
    \centering
    \includegraphics[width=\columnwidth]{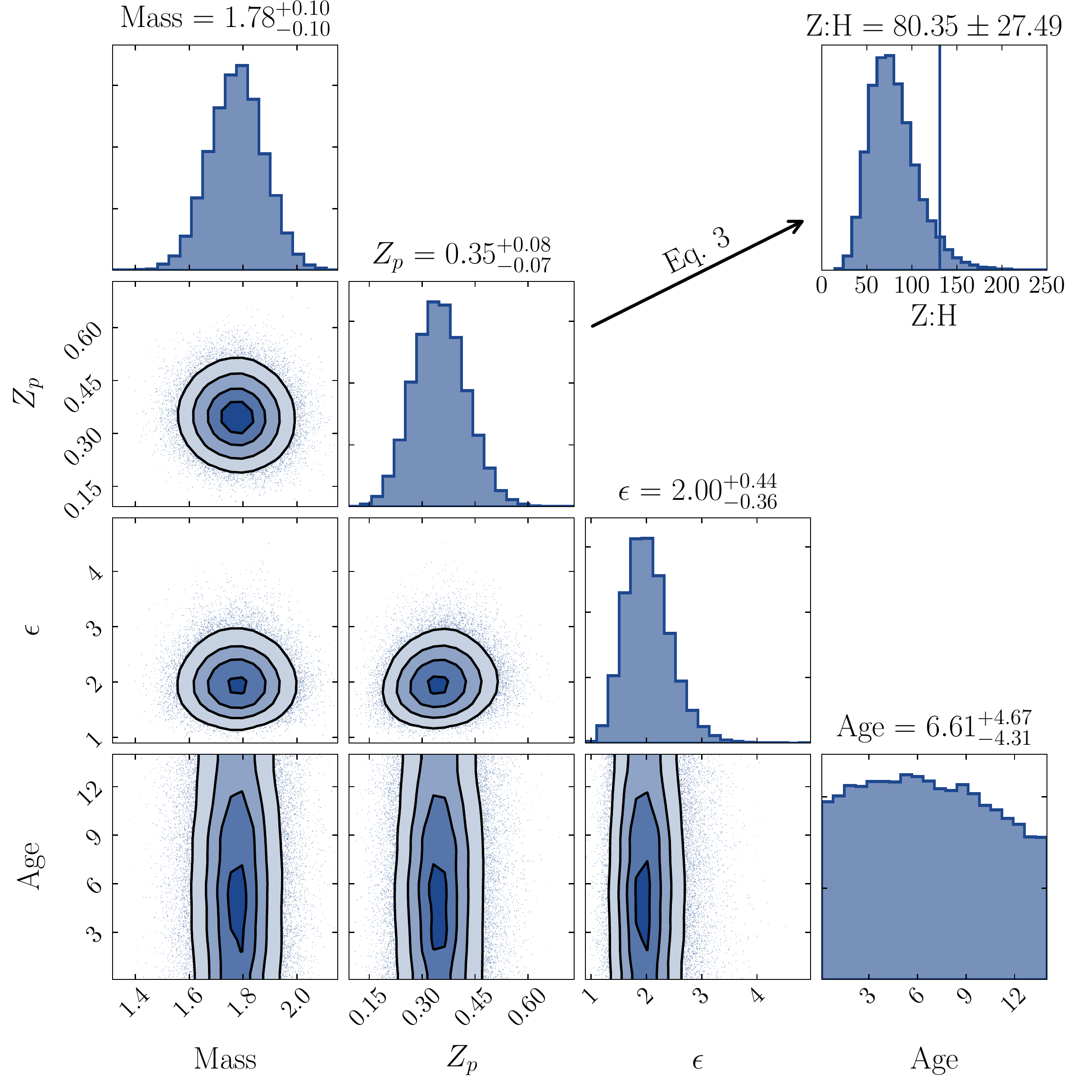}
    \caption{A corner plot of the posterior (Eq. \ref{eq:posterior}) for WASP-43 b.  The parameters are the mass of the planet in Jupiter masses, the bulk metallicity of the planet (metal fraction by mass), the anomalous heating efficiency (the fraction of incident flux deposited in the interior; see \cite{Thorngren2018}) in percent, and the age in Gyr.  A small degree of Gaussian smoothing was applied to the 2-D histograms to make them clearer.  The top right histogram shows the Z:H distribution derived from the $Z_p$ posterior using Equation \ref{eq:nz}, and the upper limit at the 95th percentile ($131\times$ solar).}
    \label{fig:examplePost}
\end{figure}

We sampled from this posterior separately for each planet using a Metropolis-Hastings sampler \citep{Hastings1970}, drawing 10,000 samples in each of four independent chains, burning in for 100,000 samples and recording only every 100$^\mathrm{th}$ sample (thinning) thereafter.  Convergence was evaluated using the Gelman-Rubin diagnostic \citep{Gelman1992} and acceptance rates, as well as visual inspection of the autocorrelation plots, trace plots, and corner plots \citep{Foreman-Mackey2013}.  As an example, Fig. \ref{fig:examplePost} depicts the posterior distribution for WASP-43b.  We can see that its metallicity is $Z_p = .35 \pm .08$, uncorrelated with other parameters because the primary source of uncertainty (in this case) is the radius measurement.  With these posterior samples in hand, we can derive a distribution for $\ratio$ from $Z_p$ using Eq. \ref{eq:nz}, assuming $f=1$, which yields $\ratio$ of $80.35 \pm 27.5 \times$ solar.  We use the 95$^\mathrm{th}$ percentile of this distribution as our upper limit, which for WASP-43b is $131\times$ solar.

\section{Results}\label{sec:results}
\subsection{Known Planets}
\begin{table*}
\centering
\begin{tabular}{lrrrrrrrrrr}
\hline 
Name & Mass & Radius & Flux & $T_\mathrm{eq}$ & a & e & Period & $Z_\mathrm{p}$ & $\ratio$ & $\ratio_{max}$\\
\hline 
HAT-P-26 b & $0.07 \pm 0.02$ & $0.63 \pm 0.04$ & $0.271$ & $1046$ & $0.0479$ & $0.12$ & $4.2345$ & $0.66 \pm 0.03$ & $281.06 \pm 39.1$ & $348.5$ \\
HD 209458 b & $0.73 \pm 0.04$ & $1.39 \pm 0.02$ & $1.061$ & $1471$ & $0.0475$ & $0.00$ & $3.5247$ & $0.16 \pm 0.02$ & $28.07 \pm 3.8$ & $34.4$ \\
WASP-12 b & $1.47 \pm 0.07$ & $1.90 \pm 0.06$ & $8.933$ & $2505$ & $0.0234$ & $0.05$ & $1.0914$ & $0.09 \pm 0.02$ & $13.70 \pm 3.8$ & $20.4$ \\
WASP-17 b & $0.78 \pm 0.23$ & $1.87 \pm 0.24$ & $1.890$ & $1699$ & $0.0515$ & $0.00$ & $3.7354$ & $0.17 \pm 0.07$ & $30.62 \pm 19.7$ & $61.4$ \\
WASP-39 b & $0.28 \pm 0.03$ & $1.27 \pm 0.04$ & $0.358$ & $1121$ & $0.0486$ & $0.00$ & $4.0553$ & $0.22 \pm 0.03$ & $40.51 \pm 8.3$ & $54.5$ \\
WASP-43 b & $1.78 \pm 0.10$ & $0.93 \pm 0.08$ & $0.821$ & $1379$ & $0.0142$ & $0.00$ & $0.8135$ & $0.35 \pm 0.07$ & $80.35 \pm 27.5$ & $130.6$ \\
WASP-52 b & $0.46 \pm 0.02$ & $1.27 \pm 0.03$ & $0.647$ & $1299$ & $0.0272$ & $0.00$ & $1.7498$ & $0.23 \pm 0.02$ & $42.43 \pm 4.8$ & $50.6$ \\
WASP-107 b & $0.12 \pm 0.01$ & $0.94 \pm 0.02$ & $0.068$ & $740$ & $0.0550$ & $0.00$ & $5.7215$ & $0.24 \pm 0.04$ & $46.55 \pm 9.3$ & $62.9$ \\
\hline
\end{tabular}

\label{resultsTable}
\caption{Planetary parameters, orbital parameters, and derived quantities for a selected subset of the planets modeled.  $T_{\mathrm{eq}}$ is the equilibrium temperature for a zero-albedo planet with full atmospheric redistribution of heat.  $Z_p$ is the bulk metal mass fraction of the planet, $\ratio_p$ is the corresponding atmosphere abundance (eq. \ref{eq:nz}) assuming a fully mixed planet ($f$ = 1), and $\ratio_{max}$ is the corresponding upper limit (the 95$^\mathrm{th}$ percentile of $\ratio$).  The full table contains 403 planets and is available for download.  Discovery and data sources -- HAT-P-26 b: \cite{Hartman2011}; HD 209458 b: \cite{Henry1999}, \cite{Southworth2010}; WASP-12 b: \cite{Hebb2009}, \cite{Collins2015}; WASP-17 b: \cite{Anderson2010}, \cite{Southworth2012}; WASP-39 b:  \cite{Faedi2011}; WASP-43 b: \cite{Hellier2011}, \cite{Gillon2012}; WASP-52 b: \cite{Hebrard2013}; WASP-107 b: \cite{Anderson2017}.}
\end{table*}

Our main results are the upper limits on atmospheric metallicity $\ratio_{max}$, a selection of which are shown in Table \ref{resultsTable} along with the input parameters we used for each planet.  The posterior means and standard deviations of $Z_p$ and $\ratio_p$ are also shown for reference.  Added caution is advisable for using the $Z_p$ values, as these distributions are more sensitive to the prior on $\epsilon$ than the upper limits are.  Nevertheless, they are reasonable estimates.

\begin{figure}
    \centering
    \includegraphics[width=\columnwidth]{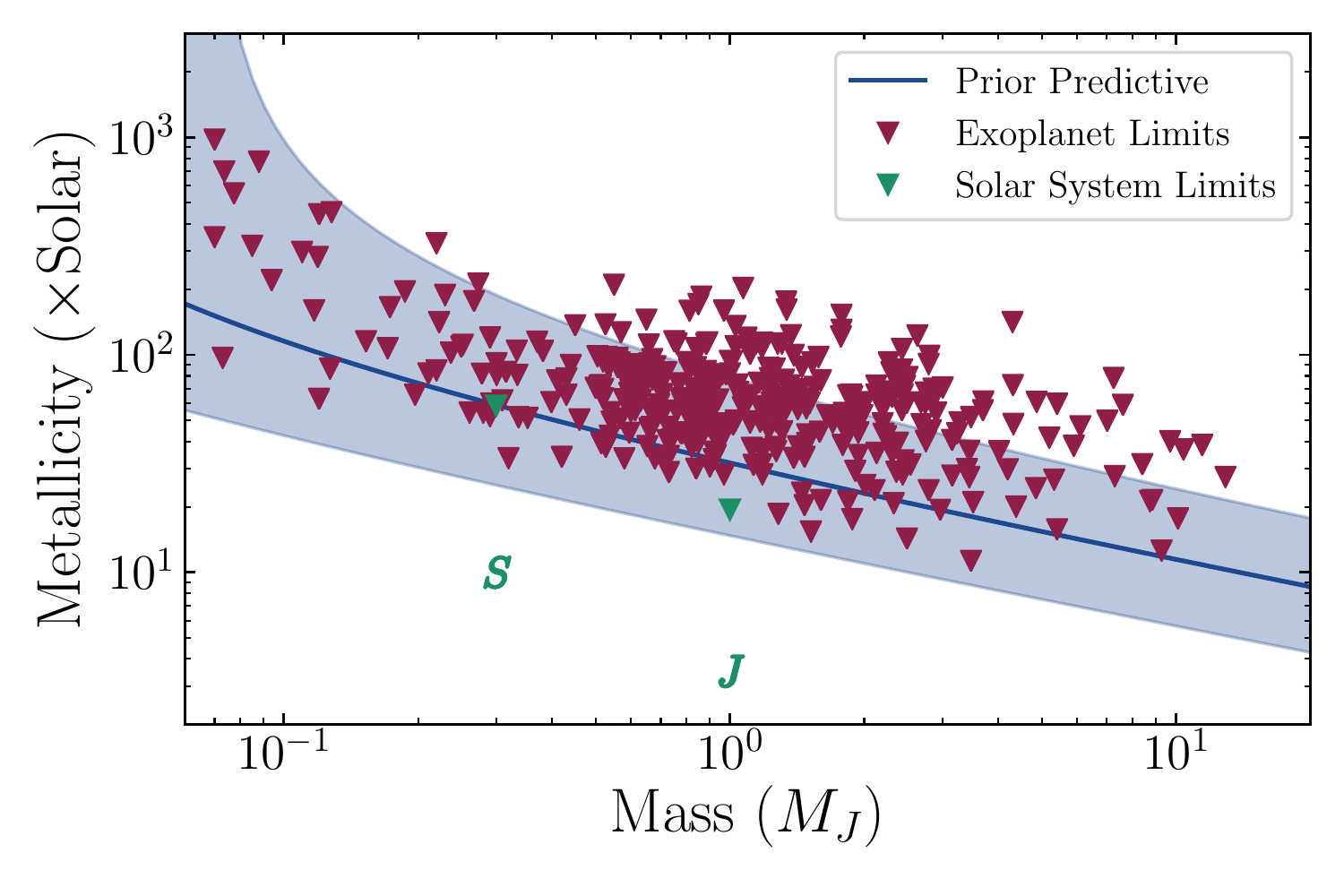}
    \caption{The computed upper limits $\ratio_{max}$ for exoplanets, Jupiter, and Saturn, plotted against mass.  Also shown is the prior predictive distribution from Eq. \ref{eq:priorPredictive}..  The limits are systematically higher than the predictive because they are the $95^\mathrm{th}$ percentile of the posterior for each planet.  The actual observed atmospheric  abundances of Jupiter and Saturn \cite{Atreya2016} are shown as \emph{J} and \emph{S}, and are about 20\% of the limits we compute.}
    \label{fig:limitsVsMass}
\end{figure}

Figure \ref{fig:limitsVsMass} shows the upper limits $\ratio_{max}$ plotted against planetary mass, along with the prior for $\ratio$ from Eq. \ref{eq:priorPredictive}.  The prior shows the expected mass-dependence of the metallicity for $f=1$, going from $\sim100 \times$ solar at Neptune masses to $<10\times$ solar for brown dwarfs.  The $1\sigma$ range for the prior is shown as a shaded region; at small masses, $Z_p$ is typically closer to the asymptote in $\ratio$ at $Z_p = 1$ (see Eq. \ref{eq:nz}), leading to larger uncertainties.  The upper limits are generally higher than the prior mean, as expected.
 
For some planets, $Z_p$ was potentially close to one.  This typically occurs for low mass planets near the cutoff of $20 M_\oplus$, or planets with larger uncertainties in mass or radius.  As $Z_p \rightarrow 1$, $\ratio \rightarrow \infty$, so we cannot provide meaningful upper-limits on $\ratio$ in that range.  To reflect this, we have identified the 21 planets whose posterior $Z_p$ has a 99$^\mathrm{th}$ percentile exceeds 0.9, and removed the upper limit.  We chose to strike the entry rather than remove the planet from the table so that readers will at least know that these planets are consistent with very large values of $\ratio$.

For comparison, we applied our models to Jupiter and Saturn.  Since these are not inflated and have tiny mass, radius, and age uncertainties, our methods produce values with negligible error bars.  Of course, for these cases, the assumption that observational error dominates modeling uncertainties \citep[discussed in][]{Thorngren2016} is not valid, but the comparison is still worth making.  For Jupiter we obtain $Z_p = .12$ and $\ratio \leq 17.7$; \cite{Guillot1999} compute $.03 \leq Z_p \leq .12$, and the observed atmospheric value is $\ratio \approx 3.5$ \citep{Atreya2016}.  For Saturn, we get $Z_p = .291$ and $\ratio \leq 51$; \cite{Guillot1999} compute compute $.21 \leq Z_p \leq .31$, and the observed atmospheric value is $\ratio \approx 10$ \citep{Atreya2016}.  In both cases, the metal abundance seen in the atmosphere is about 20\% of the value we compute for the bulk (the upper limit).  By mass, $f\approx 0.2$ (see \S \ref{sec:methods}) also.  These limits and actual values are shown in Figure \ref{fig:limitsVsMass}.

\subsection{Fits for Future Discoveries}
For new exoplanet discoveries, it would be useful to have a rough estimate of $\ratio$ in advance of running full interior structure models.  For this purpose, we have constructed least squares fits of the observed, $Z$, $\log_{10}(\ratio_p)$, $\log_{10}(\ratio_{max})$.  Due to the complexity of the underlying models, a relatively large number of predictor variables were needed; these were selected by hand with the aim of minimizing the model BIC \citep{Schwarz1978} while keeping the number of variables manageable.  The results of these fits were as follows:
\begin{align}
    \log_{10}&(Z) = \\
    &-2.02 -0.27\log_{10}(M) -4.75\log_{10}(R)+\nonumber\\
    &0.17\log_{10}(F) -1.27\log_{10}(M)\log_{10}(R)+\nonumber\\
    &0.34\log_{10}(F)\log_{10}(R)\pm 0.1\nonumber
\end{align}
\begin{align}
    \log_{10}&(\ratio) = \\
    &-0.16 -0.35\log_{10}(M) -9.32\log_{10}(R) +\nonumber\\
    &0.22\log_{10}(F) -1.17\log_{10}(M)\log_{10}(R) + \nonumber\\ &0.77\log_{10}(F)\log_{10}(R)\pm 0.13\nonumber
\end{align}
\begin{align}
    \log_{10}&(\ratio_{max}) = \\
    &1.14 -0.27\log_{10}(M) -9.06\log_{10}(R) +\nonumber\\
    &0.19\log_{10}(F) -1.15\log_{10}(M)\log_{10}(R) +\nonumber\\
    &0.72\log_{10}(F)\log_{10}(R) + 1.07\log_{10}(\sigma_R) + \nonumber\\
    &0.3\log_{10}(\sigma_R)^2\pm 0.11\nonumber
\end{align}
It is important to remember that these are only fits, and so extrapolation is not appropriate; they should only be used for planets with parameters similar to that of our data.  Using the 5 to 95$^\mathrm{th}$ percentiles, these are $.13 < M < 4.80$ ($M_J$), $.51 < R < 1.63$ ($R_J$), $.033 < F < 4.60$ (\gerg), and $.015 < \sigma_R < .22$ ($R_J$).  The other important caveat is that we have made no attempt to account for observation error, so applying these formulas to planets discovered using methods/telescopes with sensitivities significantly different than the planets we considered may produce a systematic bias.  Still, even though they are approximate, these fits provide quick and useful estimates for contextualizing new observations.

\section{Discussion}
We anticipate that these upper limits will provide useful information to atmosphere models.  For example, \cite{Wakeford2018} examine WASP-39b and find, among other results, an atmospheric metallicity of $151_{-46}^{+48}\;\times$ solar.  Our models find a maximum metallicity of $54.5\;\times$ solar.  This tension suggests that the true metallicity is near the bottom of their $2\sigma$ range, and that the planet likely has a fairly well mixed interior.  It may also point towards favoring their free chemistry model, which found a moderately lower metallicity.

In some cases, the metallicity can exceed our upper limits if the planet interior is hotter than expected by our models.  This could occur if the planet is tidally heated, as hypothesized for GJ 436b in \cite{Morley2017}, or if the planet is potentially much younger than the models \citep[see discussion in][]{Kreidberg2018}.  These potential effects would be minimal in hot Jupiters if the anomalous heating mechanism does not include a delayed cooling component \citep[see][]{Fortney2010,Spiegel2013}, as these planets must already be supplied with a massive amount of energy and would quickly reach equilibrium.

In the long run as these observations become more numerous and precise, it may be possible to investigate the ratio of the atmosphere metallicity to the bulk metallicity, $f$.  If a certain set of planets consistently exhibit $f \approx 1$ (such as how WASP-39b appears), it suggests that these planets are generally well mixed -- they have minimal cores or composition gradients.  Cases where $f$ is closer to zero, such as the solar system planets, suggest the converse.  These possibilities have been studied theoretically both in the solar system \citep[e.g.][]{Vazan2016,Moll2017,Leconte2012}, and for exoplanets \cite[e.g][]{Vazan2015,Chabrier2007}, but observational evidence has been sparse, especially for the latter case.  Using interior models in conjunction with atmosphere modeling can provide a new and unique approach to these issues.

\bibliography{bibliography}

\begin{thebibliography}{}
\expandafter\ifx\csname natexlab\endcsname\relax\def\natexlab#1{#1}\fi

\bibitem[{Akeson {et~al.}(2013)Akeson, Chen, Ciardi, Crane, Good, Harbut,
  Jackson, Kane, Laity, Leifer, Lynn, McElroy, Papin, Plavchan, Ram\'irez, Rey,
  {von Braun}, Wittman, Abajian, Ali, Beichman, Beekley, Berriman, Berukoff,
  Bryden, Chan, Groom, Lau, Payne, Regelson, Saucedo, Schmitz, Stauffer, Wyatt,
  \& Zhang}]{Akeson2013}
Akeson, R.~L., Chen, X., Ciardi, D., {et~al.} 2013, Publications of the
  Astronomical Society of the Pacific, 125, 989

\bibitem[{Anderson {et~al.}(2010)Anderson, Hellier, Gillon, Triaud, Smalley,
  Hebb, Cameron, Maxted, Queloz, West, Bentley, Enoch, Horne, Lister, Mayor,
  Parley, {F. Pepe}, Pollacco, S\'egransan, Udry, \& Wilson}]{Anderson2010}
Anderson, D.~R., Hellier, C., Gillon, M., {et~al.} 2010, ApJ, 709, 159

\bibitem[{Anderson {et~al.}(2017)Anderson, Cameron, Delrez, Doyle, Gillon,
  Hellier, Jehin, Lendl, Maxted, Madhusudhan, Pepe, Pollacco, Queloz,
  S\'egransan, Smalley, Smith, Triaud, Turner, Udry, \& West}]{Anderson2017}
Anderson, D.~R., Cameron, A.~C., Delrez, L., {et~al.} 2017, A\&A, 604, A110

\bibitem[{Asplund {et~al.}(2009)Asplund, Grevesse, Sauval, \&
  Scott}]{Asplund2009}
Asplund, M., Grevesse, N., Sauval, A.~J., \& Scott, P. 2009, Annual Review of
  Astronomy and Astrophysics, 47, 481

\bibitem[{Atreya {et~al.}(2016)Atreya, Crida, Guillot, Lunine, Madhusudhan, \&
  Mousis}]{Atreya2016}
Atreya, S.~K., Crida, A., Guillot, T., {et~al.} 2016, ArXiv e-prints, 1606,
  arXiv:1606.04510

\bibitem[{Barclay {et~al.}(2018)Barclay, Pepper, \& Quintana}]{Barclay2018}
Barclay, T., Pepper, J., \& Quintana, E.~V. 2018, arXiv:1804.05050 [astro-ph],
  arXiv:1804.05050

\bibitem[{Bean {et~al.}(2018)Bean, Stevenson, Batalha, {Berta-Thompson},
  Kreidberg, Crouzet, Benneke, Line, Sing, Wakeford, Knutson, Kempton,
  D\'esert, Crossfield, Batalha, {de Wit}, Parmentier, Harrington, Moses,
  {Lopez-Morales}, Alam, Blecic, Bruno, Carter, Chapman, Decin, Dragomir,
  Evans, Fortney, Fraine, Gao, Garc\'ia Mu\~noz, Gibson, Goyal, Heng, Hu,
  Kendrew, Kilpatrick, Krick, Lagage, Lendl, Louden, Madhusudhan, Mandell,
  Mansfield, May, Morello, Morley, Nikolov, Redfield, Roberts, Schlawin, Spake,
  Todorov, Tsiaras, Venot, Waalkes, Wheatley, Zellem, Angerhausen, Barrado,
  Carone, Casewell, Cubillos, Damiano, {de Val-Borro}, Drummond, Edwards, Endl,
  Espinoza, France, Gizis, Greene, Henning, Hong, Ingalls, Iro, Irwin, Kataria,
  Lahuis, Leconte, {Lillo-Box}, Lines, Mancini, Marchis, Mayne, Palle, Roudier,
  Shkolnik, Southworth, Teske, Tinetti, Tremblin, Tucker, {vanBoekel},
  Waldmann, Weaver, \& Zingales}]{Bean2018}
Bean, J.~L., Stevenson, K.~B., Batalha, N.~M., {et~al.} 2018, ArXiv e-prints,
  1803, arXiv:1803.04985

\bibitem[{Beichman {et~al.}(2014)Beichman, Benneke, Knutson, Smith, Lagage,
  Dressing, Latham, Lunine, Birkmann, Ferruit, Giardino, Kempton, Carey, Krick,
  Deroo, Mandell, Ressler, Shporer, Swain, Vasisht, Ricker, Bouwman,
  Crossfield, Greene, Howell, Christiansen, Ciardi, Clampin, Greenhouse,
  Sozzetti, Goudfrooij, Hines, Keyes, Lee, McCullough, Robberto, Stansberry,
  Valenti, Rieke, Rieke, Fortney, Bean, Kreidberg, Ehrenreich, Deming, Albert,
  Doyon, \& Sing}]{Beichman2014}
Beichman, C., Benneke, B., Knutson, H., {et~al.} 2014, Publications of the
  Astronomical Society of the Pacific, 126, 1134

\bibitem[{Benneke \& Seager(2012)}]{Benneke2012}
Benneke, B., \& Seager, S. 2012, The Astrophysical Journal, 753, 100

\bibitem[{Burrows {et~al.}(2008)Burrows, Budaj, \& Hubeny}]{Burrows2008}
Burrows, A., Budaj, J., \& Hubeny, I. 2008, The Astrophysical Journal, 678,
  1436

\bibitem[{Chabrier \& Baraffe(2007)}]{Chabrier2007}
Chabrier, G., \& Baraffe, I. 2007, The Astrophysical Journal Letters, 661, L81

\bibitem[{Collins {et~al.}(2015)Collins, Kielkopf, \& Stassun}]{Collins2015}
Collins, K.~A., Kielkopf, J.~F., \& Stassun, K.~G. 2015, arXiv e-prints,
  arXiv:1512.00464

\bibitem[{Demory \& Seager(2011)}]{Demory2011}
Demory, B.-O., \& Seager, S. 2011, ApJS, 197, 12

\bibitem[{Espinoza {et~al.}(2017)Espinoza, Fortney, Miguel, Thorngren, \&
  {Murray-Clay}}]{Espinoza2017}
Espinoza, N., Fortney, J.~J., Miguel, Y., Thorngren, D., \& {Murray-Clay}, R.
  2017, The Astrophysical Journal Letters, 838, L9

\bibitem[{Evans {et~al.}(2016)Evans, Sing, Wakeford, Nikolov, Ballester,
  Drummond, Kataria, Gibson, Amundsen, \& Spake}]{Evans2016}
Evans, T.~M., Sing, D.~K., Wakeford, H.~R., {et~al.} 2016, The Astrophysical
  Journal Letters, 822, L4

\bibitem[{Faedi {et~al.}(2011)Faedi, Barros, Anderson, Brown, Cameron,
  Pollacco, Boisse, H\'ebrard, Lendl, Lister, Smalley, Street, Triaud, Bento,
  Bouchy, Butters, Enoch, Haswell, Hellier, Keenan, Miller, Moulds, Moutou,
  Norton, Queloz, Santerne, Simpson, Skillen, Smith, Udry, Watson, West, \&
  Wheatley}]{Faedi2011}
Faedi, F., Barros, S. C.~C., Anderson, D.~R., {et~al.} 2011, A\&A, 531, A40

\bibitem[{Fisher \& Heng(2018)}]{Fisher2018}
Fisher, C., \& Heng, K. 2018, arXiv:1809.06894 [astro-ph, physics:physics],
  arXiv:1809.06894

\bibitem[{{Foreman-Mackey} {et~al.}(2013){Foreman-Mackey}, Hogg, Lang, \&
  Goodman}]{Foreman-Mackey2013}
{Foreman-Mackey}, D., Hogg, D.~W., Lang, D., \& Goodman, J. 2013, Publications
  of the Astronomical Society of the Pacific, 125, 306

\bibitem[{Fortney(2005)}]{Fortney2005}
Fortney, J.~J. 2005, Monthly Notices of the Royal Astronomical Society, 364,
  649

\bibitem[{Fortney {et~al.}(2008)Fortney, Lodders, Marley, \&
  Freedman}]{Fortney2008}
Fortney, J.~J., Lodders, K., Marley, M.~S., \& Freedman, R.~S. 2008, The
  Astrophysical Journal, 678, 1419

\bibitem[{Fortney {et~al.}(2013)Fortney, Mordasini, Nettelmann, Kempton,
  Greene, \& {Kevin Zahnle}}]{Fortney2013}
Fortney, J.~J., Mordasini, C., Nettelmann, N., {et~al.} 2013, ApJ, 775, 80

\bibitem[{Fortney \& Nettelmann(2010)}]{Fortney2010}
Fortney, J.~J., \& Nettelmann, N. 2010, Space Science Reviews, 152, 423

\bibitem[{French {et~al.}(2009)French, Mattsson, Nettelmann, \&
  Redmer}]{French2009}
French, M., Mattsson, T.~R., Nettelmann, N., \& Redmer, R. 2009, Phys. Rev. B,
  79, 054107

\bibitem[{Gardner {et~al.}(2006)Gardner, Mather, Clampin, Doyon, Greenhouse,
  Hammel, Hutchings, Jakobsen, Lilly, Long, Lunine, McCaughrean, Mountain,
  Nella, Rieke, Rieke, Rix, Smith, Sonneborn, Stiavelli, Stockman, Windhorst,
  \& Wright}]{Gardner2006}
Gardner, J.~P., Mather, J.~C., Clampin, M., {et~al.} 2006, Space Science
  Reviews, 123, 485

\bibitem[{Gelman \& Rubin(1992)}]{Gelman1992}
Gelman, A., \& Rubin, D.~B. 1992, Statist. Sci., 7, 457

\bibitem[{Gibson {et~al.}(2012)Gibson, Aigrain, Pont, Sing, D\'esert, Evans,
  Henry, Husnoo, \& Knutson}]{Gibson2012}
Gibson, N.~P., Aigrain, S., Pont, F., {et~al.} 2012, Monthly Notices of the
  Royal Astronomical Society, 422, 753

\bibitem[{Gillon {et~al.}(2012)Gillon, Triaud, Fortney, Demory, Jehin, Lendl,
  Magain, Kabath, Queloz, Alonso, Anderson, Cameron, Fumel, Hebb, Hellier,
  Lanotte, Maxted, Mowlavi, \& Smalley}]{Gillon2012}
Gillon, M., Triaud, A. H. M.~J., Fortney, J.~J., {et~al.} 2012, Astronomy \&
  Astrophysics, 542, A4

\bibitem[{Griffith(2014)}]{Griffith2014a}
Griffith, C.~A. 2014, Phil. Trans. R. Soc. A, 372, 20130086

\bibitem[{Guillot(1999)}]{Guillot1999}
Guillot, T. 1999, Planetary and Space Science, 47, 1183

\bibitem[{Hartman {et~al.}(2011)Hartman, Bakos, Kipping, Torres, Kov\'acs,
  Noyes, Latham, Howard, Fischer, Johnson, Marcy, Isaacson, Quinn, Buchhave,
  B\'eky, Sasselov, Stefanik, Esquerdo, Everett, Perumpilly, L\'az\'ar, Papp,
  \& S\'ari}]{Hartman2011}
Hartman, J.~D., Bakos, G.~A., Kipping, D.~M., {et~al.} 2011, The Astrophysical
  Journal, 728, 138

\bibitem[{Hastings(1970)}]{Hastings1970}
Hastings, W.~K. 1970, Biometrika, 57, 97

\bibitem[{Hebb {et~al.}(2009)Hebb, {Collier-Cameron}, Loeillet, Pollacco,
  H\'ebrard, Street, Bouchy, Stempels, Moutou, Simpson, Udry, Joshi, West,
  Skillen, Wilson, McDonald, Gibson, Aigrain, Anderson, Benn, Christian, Enoch,
  Haswell, Hellier, Horne, Irwin, Lister, Maxted, Mayor, Norton, Parley, Pont,
  Queloz, Smalley, \& Wheatley}]{Hebb2009}
Hebb, L., {Collier-Cameron}, A., Loeillet, B., {et~al.} 2009, ApJ, 693, 1920

\bibitem[{H\'ebrard {et~al.}(2013)H\'ebrard, Cameron, Brown, D\'iaz, Faedi,
  Smalley, Anderson, Armstrong, Barros, Bento, Bouchy, Doyle, Enoch, Chew,
  H\'ebrard, Hellier, Lendl, Lister, Maxted, McCormac, Moutou, Pollacco,
  Queloz, Santerne, Skillen, Southworth, {Tregloan-Reed}, Triaud, Udry,
  Vanhuysse, Watson, West, \& Wheatley}]{Hebrard2013}
H\'ebrard, G., Cameron, A.~C., Brown, D. J.~A., {et~al.} 2013, A\&A, 549, A134

\bibitem[{Hellier {et~al.}(2011)Hellier, Anderson, Collier~Cameron, Gillon,
  Jehin, Lendl, Maxted, Pepe, Pollacco, Queloz, S\'egransan, Smalley, Smith,
  Southworth, Triaud, Udry, \& West}]{Hellier2011}
Hellier, C., Anderson, D.~R., Collier~Cameron, A., {et~al.} 2011, Astronomy and
  Astrophysics, 535, L7

\bibitem[{Heng(2018)}]{Heng2018}
Heng, K. 2018, ArXiv e-prints, 1807, arXiv:1807.06102

\bibitem[{Heng \& Kitzmann(2017)}]{Heng2017}
Heng, K., \& Kitzmann, D. 2017, Monthly Notices of the Royal Astronomical
  Society, 470, 2972

\bibitem[{Henry {et~al.}(1999)Henry, Marcy, Butler, \& Vogt}]{Henry1999}
Henry, G.~W., Marcy, G.~W., Butler, R.~P., \& Vogt, S.~S. 1999, ApJ, 529, L41

\bibitem[{Huang {et~al.}(2018)Huang, Burt, Vanderburg, G\"unther, Shporer,
  Dittmann, Winn, Wittenmyer, Sha, Kane, Ricker, Vanderspek, Latham, Seager,
  Jenkins, Caldwell, Collins, Guerrero, Smith, Quinn, Udry, Pepe, Bouchy,
  {gransan}, Lovis, Ehrenreich, Marmier, Mayor, Wohler, Haworth, Morgan,
  Fausnaugh, Charbonneau, Narita, \& {team}}]{Huang2018}
Huang, C.~X., Burt, J., Vanderburg, A., {et~al.} 2018, arXiv:1809.05967
  [astro-ph], arXiv:1809.05967

\bibitem[{Knutson {et~al.}(2008)Knutson, Charbonneau, Allen, Burrows, \&
  Megeath}]{Knutson2008}
Knutson, H.~A., Charbonneau, D., Allen, L.~E., Burrows, A., \& Megeath, S.~T.
  2008, The Astrophysical Journal, 673, 526

\bibitem[{Kreidberg {et~al.}(2018)Kreidberg, Line, Thorngren, Morley, \&
  Stevenson}]{Kreidberg2018}
Kreidberg, L., Line, M.~R., Thorngren, D., Morley, C.~V., \& Stevenson, K.~B.
  2018, The Astrophysical Journal Letters, 858, L6

\bibitem[{Kreidberg {et~al.}(2014)Kreidberg, Bean, D\'esert, Benneke, Deming,
  Stevenson, Seager, {Berta-Thompson}, Seifahrt, \& Homeier}]{Kreidberg2014a}
Kreidberg, L., Bean, J.~L., D\'esert, J.-M., {et~al.} 2014, Nature, 505, 69

\bibitem[{Leconte \& Chabrier(2012)}]{Leconte2012}
Leconte, J., \& Chabrier, G. 2012, A\&A, 540, A20

\bibitem[{Line {et~al.}(2014)Line, Knutson, Wolf, \& Yung}]{Line2014}
Line, M.~R., Knutson, H., Wolf, A.~S., \& Yung, Y.~L. 2014, The Astrophysical
  Journal, 783, 70

\bibitem[{Line \& Parmentier(2016)}]{Line2016}
Line, M.~R., \& Parmentier, V. 2016, The Astrophysical Journal, 820, 78

\bibitem[{Madhusudhan {et~al.}(2014)Madhusudhan, Amin, \&
  Kennedy}]{Madhusudhan2014}
Madhusudhan, N., Amin, M.~A., \& Kennedy, G.~M. 2014, The Astrophysical Journal
  Letters, 794, L12

\bibitem[{Madhusudhan {et~al.}(2011)Madhusudhan, Harrington, Stevenson,
  Nymeyer, Campo, Wheatley, Deming, Blecic, Hardy, Lust, Anderson,
  {Collier-Cameron}, Britt, Bowman, Hebb, Hellier, Maxted, Pollacco, \&
  West}]{Madhusudhan2011}
Madhusudhan, N., Harrington, J., Stevenson, K.~B., {et~al.} 2011, Nature, 469,
  64

\bibitem[{Mandell {et~al.}(2013)Mandell, Haynes, Sinukoff, Madhusudhan,
  Burrows, \& Deming}]{Mandell2013}
Mandell, A.~M., Haynes, K., Sinukoff, E., {et~al.} 2013, The Astrophysical
  Journal, 779, 128

\bibitem[{Miller \& Fortney(2011)}]{Miller2011}
Miller, N., \& Fortney, J.~J. 2011, The Astrophysical Journal Letters, 736, L29

\bibitem[{Moll {et~al.}(2017)Moll, Garaud, Mankovich, \& Fortney}]{Moll2017}
Moll, R., Garaud, P., Mankovich, C., \& Fortney, J.~J. 2017, The Astrophysical
  Journal, 849, 24

\bibitem[{Mordasini {et~al.}(2016)Mordasini, {van Boekel}, Molli\`ere, Henning,
  \& Benneke}]{Mordasini2016}
Mordasini, C., {van Boekel}, R., Molli\`ere, P., Henning, T., \& Benneke, B.
  2016, The Astrophysical Journal, 832, 41

\bibitem[{Morley {et~al.}(2013)Morley, Fortney, Kempton, Marley, Visscher, \&
  Zahnle}]{Morley2013}
Morley, C.~V., Fortney, J.~J., Kempton, E. M.-R., {et~al.} 2013, The
  Astrophysical Journal, 775, 33

\bibitem[{Morley {et~al.}(2017)Morley, Knutson, Line, Fortney, Thorngren,
  Marley, Teal, \& Lupu}]{Morley2017}
Morley, C.~V., Knutson, H., Line, M., {et~al.} 2017, The Astronomical Journal,
  153, 86

\bibitem[{\"Oberg {et~al.}(2011)\"Oberg, {Murray-Clay}, \& Bergin}]{Oberg2011a}
\"Oberg, K.~I., {Murray-Clay}, R., \& Bergin, E.~A. 2011, The Astrophysical
  Journal Letters, 743, L16

\bibitem[{Oreshenko {et~al.}(2017)Oreshenko, Lavie, Grimm, Tsai, Malik, Demory,
  Mordasini, Alibert, Benz, Quanz, Trotta, \& Heng}]{Oreshenko2017}
Oreshenko, M., Lavie, B., Grimm, S.~L., {et~al.} 2017, The Astrophysical
  Journal Letters, 847, L3

\bibitem[{Pollack {et~al.}(1996)Pollack, Hubickyj, Bodenheimer, Lissauer,
  Podolak, \& Greenzweig}]{Pollack1996}
Pollack, J.~B., Hubickyj, O., Bodenheimer, P., {et~al.} 1996, Icarus, 124, 62

\bibitem[{Ricker {et~al.}(2015)Ricker, Winn, Vanderspek, Latham, Bakos, Bean,
  {Berta-Thompson}, Brown, Buchhave, Butler, Butler, Chaplin, Charbonneau,
  {Christensen-Dalsgaard}, Clampin, Deming, Doty, De~Lee, Dressing, Dunham,
  Endl, Fressin, Ge, Henning, Holman, Howard, Ida, Jenkins, Jernigan, Johnson,
  Kaltenegger, Kawai, Kjeldsen, Laughlin, Levine, Lin, Lissauer, MacQueen,
  Marcy, McCullough, Morton, Narita, Paegert, Palle, Pepe, Pepper, Quirrenbach,
  Rinehart, Sasselov, Sato, Seager, Sozzetti, Stassun, Sullivan, Szentgyorgyi,
  Torres, Udry, \& Villasenor}]{Ricker2015}
Ricker, G.~R., Winn, J.~N., Vanderspek, R., {et~al.} 2015, Journal of
  Astronomical Telescopes, Instruments, and Systems, 1, 014003

\bibitem[{Schneider {et~al.}(2011)Schneider, Dedieu, Le~Sidaner, Savalle, \&
  Zolotukhin}]{Schneider2011}
Schneider, J., Dedieu, C., Le~Sidaner, P., Savalle, R., \& Zolotukhin, I. 2011,
  Astronomy and Astrophysics, 532, A79

\bibitem[{Schwarz(1978)}]{Schwarz1978}
Schwarz, G. 1978, Ann. Statist., 6, 461

\bibitem[{Sing {et~al.}(2011)Sing, Pont, Aigrain, Charbonneau, D\'esert,
  Gibson, Gilliland, Hayek, Henry, Knutson, Lecavelier Des~Etangs, Mazeh, \&
  Shporer}]{Sing2011}
Sing, D.~K., Pont, F., Aigrain, S., {et~al.} 2011, Monthly Notices of the Royal
  Astronomical Society, 416, 1443

\bibitem[{Southworth(2010)}]{Southworth2010}
Southworth, J. 2010, Monthly Notices of the Royal Astronomical Society, 408,
  1689

\bibitem[{Southworth {et~al.}(2012)Southworth, Hinse, Dominik, Fang,
  Harps\o{}e, J\o{}rgensen, Kerins, Liebig, Mancini, Skottfelt, Anderson,
  Smalley, {Tregloan-Reed}, Wertz, Alsubai, Bozza, Calchi~Novati, Dreizler, Gu,
  Hundertmark, {Jessen-Hansen}, Kains, Kjeldsen, Lund, Lundkvist, Mathiasen,
  Penny, Rahvar, Ricci, Scarpetta, Snodgrass, \& Surdej}]{Southworth2012}
Southworth, J., Hinse, T.~C., Dominik, M., {et~al.} 2012, Monthly Notices of
  the Royal Astronomical Society, 426, 1338

\bibitem[{Spiegel \& Burrows(2013)}]{Spiegel2013}
Spiegel, D.~S., \& Burrows, A. 2013, The Astrophysical Journal, 772, 76

\bibitem[{Sullivan {et~al.}(2015)Sullivan, Winn, {Berta-Thompson}, Charbonneau,
  Deming, Dressing, Latham, Levine, McCullough, Morton, Ricker, Vanderspek, \&
  Woods}]{Sullivan2015}
Sullivan, P.~W., Winn, J.~N., {Berta-Thompson}, Z.~K., {et~al.} 2015, The
  Astrophysical Journal, 809, 77

\bibitem[{Swain {et~al.}(2010)Swain, Deroo, Griffith, Tinetti, Thatte, Vasisht,
  Chen, Bouwman, Crossfield, Angerhausen, Afonso, \& Henning}]{Swain2010}
Swain, M.~R., Deroo, P., Griffith, C.~A., {et~al.} 2010, Nature, 463, 637

\bibitem[{Thompson(1990)}]{Thompson1990}
Thompson, S. 1990, {{ANEOS Analytic Equations}} of {{State}} for {{Shock
  Physics Codes Input Manual}},  {Sandia National Laboratory}

\bibitem[{Thorngren \& Fortney(2018)}]{Thorngren2018}
Thorngren, D.~P., \& Fortney, J.~J. 2018, The Astronomical Journal, 155, 214

\bibitem[{Thorngren {et~al.}(2016)Thorngren, Fortney, {Murray-Clay}, \&
  Lopez}]{Thorngren2016}
Thorngren, D.~P., Fortney, J.~J., {Murray-Clay}, R.~A., \& Lopez, E.~D. 2016,
  The Astrophysical Journal, 831, 64

\bibitem[{Vazan {et~al.}(2015)Vazan, Helled, Kovetz, \& Podolak}]{Vazan2015}
Vazan, A., Helled, R., Kovetz, A., \& Podolak, M. 2015, The Astrophysical
  Journal, 803, 32

\bibitem[{Vazan {et~al.}(2016)Vazan, Helled, Podolak, \& Kovetz}]{Vazan2016}
Vazan, A., Helled, R., Podolak, M., \& Kovetz, A. 2016, The Astrophysical
  Journal, 829, 118

\bibitem[{Wakeford {et~al.}(2018)Wakeford, Sing, Deming, Lewis, Goyal, Wilson,
  Barstow, Kataria, Drummond, Evans, Carter, Nikolov, Knutson, Ballester, \&
  Mandell}]{Wakeford2018}
Wakeford, H.~R., Sing, D.~K., Deming, D., {et~al.} 2018, The Astronomical
  Journal, 155, 29

\end{thebibliography}
\end{document}